\documentclass[preprint,review]{elsarticle}

 \usepackage{graphics}
\usepackage[update]{epstopdf}

\usepackage{amsmath,amssymb,amsfonts,amsthm,amsbsy,amstext}
\usepackage{siunitx}		
\usepackage{multirow} 	
\usepackage{booktabs} 	

\biboptions{square,comma,sort&compress}

\journal{Corrosion Science}

\begin{document}

\begin{frontmatter}



\title{Hydrogen accommodation in Zr second phase particles: Implications for H pick-up and hydriding of Zircaloy-2 and Zircaloy-4.}


\author[IC,ANSTO]{P.A. Burr}
\author[IC]{S.T. Murphy}
\author[IC,HMS]{S.C. Lumley}
\author[IC]{M.R. Wenman\corref{Wenman}}
\author[IC]{R.W. Grimes}
\address[IC]{Centre for Nuclear Engineering and Department of Materials, Imperial College London, London, SW7 2AZ, UK.}
\address[ANSTO]{Institute of Materials Engineering, Australian Nuclear Science \& Technology Organisation, Menai, New South Wales 2234, Australia.}
\address[HMS]{Nuclear Department, Defence Academy, HMS Sultan, Gosport, Hampshire, PO12 3BY, UK.}
\cortext[Wenman]{Corresponding author. Tel: +44 (0)2075946763\\ \indent \indent E-mail address: m.wenman@imperial.ac.uk}

\begin{abstract}
{\it Ab-initio} computer simulations have been used to predict the energies associated with the accommodation of H atoms at interstitial sites in $\alpha$, $\beta$-Zr and Zr-M intermetallics formed with common alloying additions (M = Cr, Fe, Ni).
Intermetallics that relate to the Zr$_2$(Ni,Fe) second phase particles (SPPs) found in Zircaloy-2 exhibit favourable solution enthalpies for H. The intermetallic phases that relate to the Zr(Cr,Fe)$_2$ SPPs, found predominantly in Zircaloy-4, do not offer favourable sites for interstitial H.
It is proposed that Zr(Cr,Fe)$_2$ particles may act as bridges for the migration of H through the oxide layer, whilst the Zr$_2$(Ni,Fe)-type particles will trap the migrating H until these are dissolved or fully oxidised.

\end{abstract}


\end{frontmatter}


\section{Introduction}
For the past five decades, Sn-containing Zr alloys, such as Zircaloys, have been widely used for nuclear fuel cladding and internal components of light water nuclear reactors\cite{Sabol2005}.
To improve their corrosion resistance, Zircaloy-2 (or Zry-2) also contain small amounts of Fe, Cr and Ni, which precipitate out as SPPs due to their low solid solubility in $\alpha$-Zr. 

Later alloy development led to the formulation of Zry-4, an alloy with much lower Ni content.
The reduction of Ni content in Zry-4 resulted in a reduced H pick-up fraction (HPUF) of the alloy without compromising the oxidation resistance\cite{Sabol2005}.
However, the mechanism through which the HPUF is affected by the presence of Ni is still not well understood. There is, nevertheless, general agreement that SPP composition, size and morphology play a key role\cite{Yao2011}.


There are many theories as to how intermetallics influence corrosion and H pickup but all are subject to challenge\cite{Hatano1996a,Kakiuchi2005,Bossis2005,Murai2000,Lelievre1998}.
Hatano {\it et al.}\cite{Hatano1996a}~suggested that as the oxide layer thickens, the larger SPPs residing at the oxide-metal interface will oxidise more slowly than the surrounding Zr.
These partially metallic particles could then act as a H migration pathway (bridge) through the oxide layer.
It has been noted by Shaltiel {\it et al.}\cite{Shaltiel1977}~that Laves-phase SPPs of the Zr(Fe,Cr)$_2$ type have a tendency to absorb H.  Furthermore, the ratio of Fe to Cr seems to influence H absorption. As SPPs are irradiated the Fe dissolves out first, therefore the ratio of Fe-Cr will change and concomitantly the capacity for H absorption.  This may lead to a release of H into the {$\alpha$}-Zr matrix phase late in fuel life, which could accelerate hydriding of the cladding. Equally, subsequent annealing of SPPs, heated up during the early stage of dry storage of fuel, could reabsorb H from the matrix into recrystallized SPPs and hence be beneficial in reducing available H for hydride reorientation.  At present there is little data regarding which traps for H are most efficient.

In this paper we employed density functional theory (DFT) to calculate the solution enthalpies of H in the various binary intermetallic SPPs and compares these to the solution enthalpies of H in both $\alpha$ and $\beta$-Zr phases. 
The two most common SPPs are the ternary Zr(Cr,Fe)$_2$ (especially in Zry-4) and Zr$_2$(Fe,Ni) (in Zry-2).
Whilst it has been reported that binary phases do not tend to form in Zr alloys \cite{Chemelle1983,Yang1986,Bangaru1985}, an investigation of the simple binary systems is important, in the first instance, to understand the role that individual alloying elements play in the interaction between H and the intermetallic phases.
Consequently, this allows us to determine which SPPs are likely to behave as described by Hatano's model, if at all.

\begin{figure}[p]
\includegraphics[width = \textwidth]{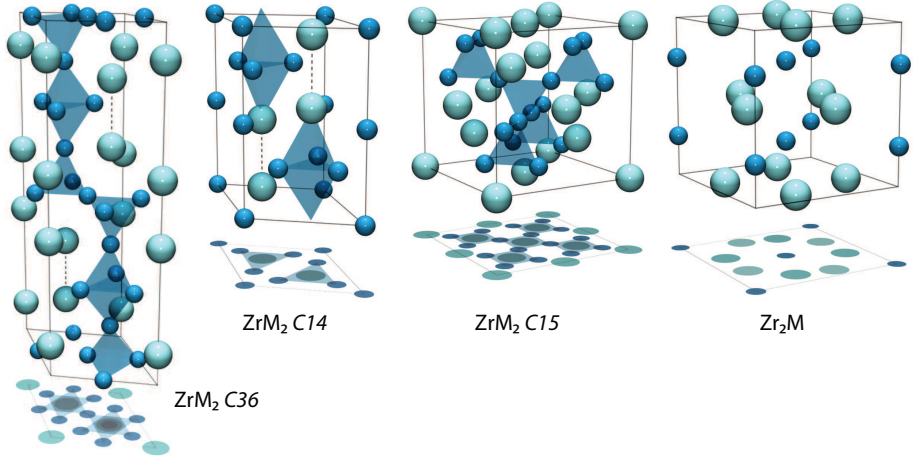}
\caption{\label{fig:ZrIM-Structures} Unit cells of {\it C36}, {\it C14}, {\it C15} Laves structure and of Zr$_2$M. The larger lighter spheres represent the Zr atoms, the smaller darker spheres represent Cr or Fe in the laves phases and Fe or Ni in the Zr-rich phase.}
\end{figure}

Cr forms three intermetallic phases with Zr: the cubic $\alpha$ phase and the hexagonal $\beta$ and $\gamma$ phases. All three have the same stoichiometric formula ZrCr$_2$ and are all Laves phases, termed {\it C15}, {\it C36} and {\it C14} respectively (see Table~\ref{tab:SuperCells} for further details).
Even though the stable phase at reactor operating temperature is $\alpha$-ZrCr$_2$, there have been many reports of both cubic\cite{Yang1987,Vitikainen1978,Versaci1979,Krasevec1981,Kuwae1983a} and hexagonal\cite{Vandersande1974,Versaci1979,Krasevec1981,Versaci1983,Kuwae1983a,Chemelle1983,Bangaru1985,Yang1986,Lelievre2002} structures in Zr alloys, therefore all three have been considered here.
Although the Zr-Ni binary system includes numerous intermetallic phases\cite{Okamoto2007}, Zr-Ni SPPs tend to be stable as Zr rich phases, in particular tetragonal Zr$_2$Ni.
Fe is found in the ZrCr$_2$-type SPPs as well as in the Zr$_2$Ni-type, therefore the {\it C15}, {\it C36} and {\it C14} ZrFe$_2$ Laves phase and the Zr$_2$Fe phase have all been studied.
Figure~\ref{fig:ZrIM-Structures} contains schematic representation of unit cells of each of the intermetallic phases described above.

\section{Computational Methodology}
All DFT simulations were carried out using CASTEP 5.5\cite{Clark2005}. Ultra-soft pseudo potentials with a consistent cut-off energy of \SI{450}{\electronvolt} were used throughout.
Previous work by Domain {\it et al.}\cite{Domain2002a}~demonstrated that the Generalized Gradient Approximation (GGA) is better suited than the Local Density Approximation to describe the exchange-correlation functional of solid Zr. Therefore, the Perdew Burke and Ernzerhof (PBE)\cite{Perdew1996} parametrisation of the GGA was adopted for this study.
A high density of {\bf k}-points was employed for the integration of the Brillouin Zone, following the  Monkhost-Pack sampling scheme\cite{Monkhorst1976}: the distance between sampling points was maintained as close as possible to \SI{0.030}{\angstrom^{-1}} and never above \SI{0.035}{\angstrom^{-1}}.
The fast fourier transform grid was set to be twice as dense as that of the wavefunctions, with a finer grid for augmentation charges scaled by 2.3.
Due to the metallic nature of the system, density mixing and 
Methfessel-Paxton\cite{Methfessel1989} cold smearing of bands were employed with a width of \SI{0.1}{eV}. Testing was carried out to ensure a convergence of \SI{e-3}{eV/atom} was achieved with respect to all of the above parameters. All calculations were spin polarised and no symmetry operations were enforced.

The energy convergence criterion for self-consistent calculations was set to \SI{1e-6}{\electronvolt}. 
Similarly robust criteria were imposed for ionic energy minimisation: energy difference $<$~\SI{1e-5}{\electronvolt}, forces on individual atoms $<$~\SI{ 0.01}{\electronvolt\per\angstrom} and stress component on cell $<$~\SI{0.05}{\giga\pascal} (for constant pressure simulations).

All non-defective structures were relaxed at constant pressure (both the crystal's lattice parameters and ion positions within the supercells were subject to energy minimisation). 
Relaxed structures were then employed for simulations containing H interstitial defects. Defect simulations were performed both at constant volume (cell parameters constrained to preserve the perfect supercell's shape and volume) and at constant pressure; these replicate dilute and alloying conditions respectively.

\section{Results and Discussion}
\subsection{$\alpha$ and $\beta$-Zr}

When considering the effective H concentration it is important to bear in mind that the solubility of H in  $\alpha$-Zr is 50--60~{\small{\it wt}~ppm} at the operating temperature of a pressurised water reactor of \SI{300}{\celsius}\cite{Okamoto2006,Kearns1967}.
For $\alpha$-Zr, calculations with a simple H interstitial defect were performed for supercells containing up to 150 atoms ($5\times5\times3$); yielding an effective defect concentration of  73.7~{\it wt}~ppm~H, therefore only slightly above the reported solubility limit.
For $\beta$-Zr the solubility limit of H is orders of magnitude larger\cite{Okamoto2006}.

The lattice parameters and enthalpy of formation of each phase investigated were presented in previous work\cite{Lumley2012} and are in agreement with values available in the literature.
In addition, the dissociation energy and dimer length of the H$_2$ molecule were calculated to be \SI{4.53}{\electronvolt} and \SI{752}{\pico \meter} respectively, in excellent agreement with experimental values, \SI{4.48}{\electronvolt} and \SI{746}{\pico \meter}\cite{CRC90}.

$\alpha$-Zr presents five sites for interstitial occupancy: a tetrahedral site in Wyckoff position\cite{Hahn1993} $4f$ (with $z=\tfrac{5}{8}$), an octahedral site ($2a$), an hexahedral site ($2d$), a basal~trigonal site ($2b$), and a non-basal~trigonal site ($12k$, with $x=\tfrac{4}{9}$, $z=\tfrac{5}{12}$) --- the latter three were found to be metastable for H occupancy (i.e. upon energy minimization the H atom relaxes into either the tetrahedral or octahedral sites).
$\beta$-Zr has stable tetrahedral $12d$ and octahedral $6b$ interstitial sites. A trigonal interstitial site also exists ($24h$, with $y=\tfrac{1}{3}$), however, this too was found to be unstable for H occupancy. 

The enthalpy of solution of an isolated H atom in bulk metal, $E_{\textrm{H}}^{sol}$, according to reaction~\ref{Eq:Hreaction}, was calculated using equation~\ref{Eq:Ereaction},
\begin{gather}
\label{Eq:Hreaction}	\mathrm{ M_{x(s)} + \tfrac{1}{2} H_{2(g)} \rightarrow M_xH_{(s)} }\\
\label{Eq:Ereaction}
E^{sol}_\mathrm{H} = E^\text{\sc dft}_\mathrm{M_xH_{(s)}} - E^\text{\sc dft}_\mathrm{M_{x(s)}} - \tfrac{1}{2} E^\text{\sc dft}_\mathrm{H_{2(g)}} 
\end{gather}
where M is a metal simulated with a supercell containing x atoms.

The calculated values of solution enthalpies for interstitial H in pure Zr are presented in Table~\ref{tab:Esol-Zr}, together with reference values from both experimental and other {\it ab-initio} simulations.

In $\alpha$-Zr, it was found that H exhibits a slight preference of the tetrahedral site over the octahedral site by  \SI{-0.086}{eV}, however, given the small energy difference between the two sites, H interstitial atom will probably occupy both sites at reactor temperatures.
Previous studies by Domain {\it et al.}\cite{Domain2002a}~record a similar relative preference for the tetrahedral site over the octahedral site, with a similar energy difference, however, their values were lower by \SI{0.13}{\electronvolt}. This discrepancy is mostly accounted for by the calculation of the H$_2$ molecule: the dissociation energy calculated by Domain {\it et al.}\cite{Domain2002a}~was \SI{0.26}{\electronvolt} smaller per hydrogen dimer.

In $\beta$-Zr, a the same preference for tetrahedral site was observed, but by a larger margin: \SI{-0.21}{eV}.
Because octahedral sites in $\beta$-Zr are anisotropic (slightly compressed in one dimension), the periodic repetition of such defects causes the structure to deform tetragonally.
In order to remove this computational artefact, these simulations were carried out with constraints on the cell such that a change in volume was allowed but without any change in cell shape ($a=b=c$ and $\alpha=\beta=\gamma=\SI{90}{\degree}$ at all times).

\subsection{Intermetallics of Zr}

When comparing defect energies across different phases, ideally the supercells of all solids investigated should have the same number of atoms to ensure that the defect concentrations are identical. However, differences in crystal structures means that this is not possible.
In order to minimise interactions between the defects and their replicas, large supercells were adopted, emulating a dilute H concentration in the materials, and all supercells were chosen to be as regular as possible (i.e. the total lengths of the supercells in the {\bf \emph{x}}, {\bf \emph{y}} and {\bf \emph{z}} directions are roughly equal, see Table~\ref{tab:SuperCells}).

Starting from the relaxed structure of a perfect intermetallic phase, a series of simulations were carried out, each with an H atom placed at one of the possible interstitial symmetry sites. This was repeated for every intermetallic phase; in each case the lowest energy site is reported in Table~\ref{tab:Esol_summary}.
It was found that the lowest energy sites, across all intermetallic phases, were always tetrahedral; and particularly favourable were the ones with a larger Zr/M ratio of neighbouring atoms.
The difference in solution enthalpy for H between a constant pressure and a constant volume simulation for a given H defect was consistently below \SI{e-2}{eV}, therefore the latter are omitted from Table~\ref{tab:Esol_summary}.
Finally, the table contains a comparison of solution enthalpies of H in the various intermetallics relative to $\alpha$-Zr. A comparison with $\beta$-Zr was considered less useful as an earlier study\cite{Lumley2012} showed that the intermetallics are soluble in the $\beta$ phase.

The ZrCr$_2$ and ZrFe$_2$ Laves phases that relate to the most common SPP found in Zry-4, offer unfavourable sites for H interstitial defects, compared to $\alpha$ and $\beta$-Zr.
In fact, the enthalpy of solution in ZrFe$_2$ is actually positive.
Since none of the Laves phase intermetallics for either Cr or Fe retain H compared to $\alpha$-Zr, it seems unlikely Zr(Cr,Fe)$_2$ will offer any more favourable sites. This finding is consistent with the H-uptake model of Hatano {\it et al.}\cite{Hatano1996a}, in that this type of SPP may act as a bridge for H diffusion through the oxide layer, provided it retains a partially metallic form in the surrounding ZrO$_2$, as suggested by Leli\`{e}vre {\it et al.}\cite{Lelievre1998}.
Conversely, Zr$_2$Fe and Zr$_2$Ni, which relate to the Zr-rich SPPs found in Zry-2, provide sites with low H solution enthalpy: Zr$_2$Fe similar to $\alpha$-Zr but Zr$_2$Ni significantly lower than either $\alpha$ or $\beta$-Zr.
Thus, if the solid solution Zr$_2$(Fe,Ni) was to behaves similarly to its binary end members Zr$_2$Fe or Zr$_2$Ni, these SPPs would preferentially retain H, which, in turn, would retard the transport of H into the Zr metal through the oxide layer, as suggested by Lim {\it et al.}\cite{Lim2003}. These particles would not contribute to the migration of H in the bridging SPPs model.

Finally, the current results suggest that Zr$_2$Ni-like intermetallic particles may also act as sinks for H atoms when these have reached the Zr metal, provided there is no hindrance in the migration process of H at the interface between the Zr matrix and the SPPs. Consequently, if these particles are dissolved late on in the  life of the fuel (as a result of radiation damage or thermal spikes) the trapped H may be released again into the Zr matrix.

In the current study we have considered the enthalpy of solution of H in those binary intermetallics that relate to the most common SPPs found in Zircaloy. These SPPs are reported to be ternary solid solutions\cite{Chemelle1983,Yang1986,Bangaru1985}, and in ternary compounds the enthalpy of solution may differ slightly due to electronic and elastic interaction of H with the solutes. This is the subject of further work and beyond the scope of the current letter.

\section{Conclusions}
The enthalpy of solution of H was calculated for $\alpha$, $\beta$-Zr and the Zr-Cr, Zr-Fe and Zr-Ni intermetallic phases that are relevant to Zry-2 and Zry-4 nuclear fuel cladding.

In Zr (both $\alpha$ and $\beta$) the tetrahedral site exhibits a slightly more favourable solution enthalpy for interstitial H compared to the octahedral site. A similar trend is found in each of the intermetallic phases investigated, but in addition, the sites with the largest fraction of neighbouring Zr atoms exhibit the lowest solution enthalpy.

The current study highlights that ZrCr$_2$ and ZrFe$_2$, the boundary phases of the Zr(Cr,Fe)$_2$ solid solution are unfavourable for H accommodation compared to $\alpha$-Zr.
Conversely, the end members of the Zr$_2$(Fe,Ni) solid solution accommodate hydrogen readily, even compared to $\alpha$-Zr.
Thus, following Hatano's model of H migration through the cladding oxide layer, we propose that the presence of partially metallic Zr(Cr,Fe)$_2$ SPPs will aid the transport of H through the oxide, while Zr$_2$(Fe,Ni) SPPs are not helpful, trapping the migrating H until these SPPs are dissolved or oxidised.
Therefore, the reason for the decrease in HPUF with decreasing Ni content in Zry-4 must be explained using models other than the one first proposed by Hatano.

\section{Acknowledgements}
We would like to thank the EPSRC, ANSTO and the UK-MOD for financial support. We also acknowledge Imperial College HPC for the use of resources.
M.R.\ Wenman acknowledges support from EDF Energy through a Fellowship scheme.





\begin{thebibliography}{37}
\expandafter\ifx\csname natexlab\endcsname\relax\def\natexlab#1{#1}\fi
\providecommand{\bibinfo}[2]{#2}
\ifx\xfnm\relax \def\xfnm[#1]{\unskip,\space#1}\fi
\bibitem[{Sabol(2005)}]{Sabol2005}
\bibinfo{author}{G.~P. Sabol}, in: \bibinfo{editor}{P.~Rudling},
  \bibinfo{editor}{B.~Kammenzind} (Eds.), \bibinfo{booktitle}{Zirconium in the
  Nuclear Industry: 14th International Symposium}, volume
  \bibinfo{volume}{1467} of \textit{\bibinfo{series}{ASTM-STP}}, pp.
  \bibinfo{pages}{3--24}.
\bibitem[{Okamoto(2010)}]{Okamoto2010}
\bibinfo{author}{H.~Okamoto}, \bibinfo{journal}{J. Phase Equilib. Diffus.}
  \bibinfo{volume}{31} (\bibinfo{year}{2010}) \bibinfo{pages}{411--412}.
\bibitem[{Yao et~al.(2011)Yao, Wang, Peng, Zhou, and Li}]{Yao2011}
\bibinfo{author}{M.~Yao}, \bibinfo{author}{J.~Wang}, \bibinfo{author}{J.~Peng},
  \bibinfo{author}{B.~Zhou}, \bibinfo{author}{Q.~Li}, in:
  \bibinfo{editor}{P.~Barberis} (Ed.), \bibinfo{booktitle}{Zirconium in the
  Nuclear Industry: 16th International Symposiumy}, volume
  \bibinfo{volume}{1529} of \textit{\bibinfo{series}{ASTM-STP}}, pp.
  \bibinfo{pages}{466--495}.
\bibitem[{Hatano et~al.(1996)Hatano, Isobe, Hitaka, and Sugisaki}]{Hatano1996a}
\bibinfo{author}{Y.~Hatano}, \bibinfo{author}{K.~Isobe},
  \bibinfo{author}{R.~Hitaka}, \bibinfo{author}{M.~Sugisaki},
  \bibinfo{journal}{J. Nucl. Sci. Technol.} \bibinfo{volume}{33}
  (\bibinfo{year}{1996}) \bibinfo{pages}{944--949}.
\bibitem[{Kakiuchi et~al.(2005)Kakiuchi, Itagaki, Furuya, Miyazaki, Ishii,
  Suzuki, Terai, and Yamawaki}]{Kakiuchi2005}
\bibinfo{author}{K.~Kakiuchi}, \bibinfo{author}{N.~Itagaki},
  \bibinfo{author}{T.~Furuya}, \bibinfo{author}{A.~Miyazaki},
  \bibinfo{author}{Y.~Ishii}, \bibinfo{author}{S.~Suzuki},
  \bibinfo{author}{T.~Terai}, \bibinfo{author}{M.~Yamawaki}, in:
  \bibinfo{editor}{P.~Rudling}, \bibinfo{editor}{B.~Kammenzind} (Eds.),
  \bibinfo{booktitle}{Zirconium in the Nuclear Industry: 14th International
  Symposium}, volume \bibinfo{volume}{1467} of
  \textit{\bibinfo{series}{ASTM-STP}}, pp. \bibinfo{pages}{349--366}.
\bibitem[{Bossis et~al.(2005)Bossis, Pecheur, Hanifi, Thomazet, and
  Blat}]{Bossis2005}
\bibinfo{author}{P.~Bossis}, \bibinfo{author}{D.~Pecheur},
  \bibinfo{author}{K.~Hanifi}, \bibinfo{author}{J.~Thomazet},
  \bibinfo{author}{M.~Blat}, in: \bibinfo{editor}{P.~Rudling},
  \bibinfo{editor}{B.~Kammenzind} (Eds.), \bibinfo{booktitle}{Zirconium in the
  Nuclear Industry: 14th International Symposium}, volume
  \bibinfo{volume}{1467} of \textit{\bibinfo{series}{ASTM-STP}}, pp.
  \bibinfo{pages}{494--525}.
\bibitem[{Murai et~al.(2000)Murai, Isobe, Takizawa, Mae, Cox, Maguire, and
  Motta}]{Murai2000}
\bibinfo{author}{T.~Murai}, \bibinfo{author}{T.~Isobe},
  \bibinfo{author}{Y.~Takizawa}, \bibinfo{author}{Y.~Mae},
  \bibinfo{author}{B.~Cox}, \bibinfo{author}{M.~Maguire},
  \bibinfo{author}{A.~T. Motta}, in: \bibinfo{editor}{G.~Saboll},
  \bibinfo{editor}{C.~Moan} (Eds.), \bibinfo{booktitle}{Zirconium in the
  Nuclear Industry: 12th International Symposium}, volume
  \bibinfo{volume}{1354} of \textit{\bibinfo{series}{ASTM-STP}}, pp.
  \bibinfo{pages}{623--640}.
\bibitem[{Leli\`{e}vre et~al.(1998)Leli\`{e}vre, Tessier, Iltis, Berthier, and
  Lefebvre}]{Lelievre1998}
\bibinfo{author}{G.~Leli\`{e}vre}, \bibinfo{author}{C.~Tessier},
  \bibinfo{author}{X.~Iltis}, \bibinfo{author}{B.~Berthier},
  \bibinfo{author}{F.~Lefebvre}, \bibinfo{journal}{J. Alloys Compd.}
  \bibinfo{volume}{268} (\bibinfo{year}{1998}) \bibinfo{pages}{308--317}.
\bibitem[{Shaltiel et~al.(1977)Shaltiel, Jacob, and Davidov}]{Shaltiel1977}
\bibinfo{author}{D.~Shaltiel}, \bibinfo{author}{I.~Jacob},
  \bibinfo{author}{D.~Davidov}, \bibinfo{journal}{J. Less-Common Met.}
  \bibinfo{volume}{53} (\bibinfo{year}{1977}) \bibinfo{pages}{117--131}.
\bibitem[{Chemelle et~al.(1983)Chemelle, Knorr, {Van Der Sande}, and
  Pelloux}]{Chemelle1983}
\bibinfo{author}{P.~Chemelle}, \bibinfo{author}{D.~Knorr},
  \bibinfo{author}{J.~{Van Der Sande}}, \bibinfo{author}{R.~Pelloux},
  \bibinfo{journal}{J. Nucl. Mater.} \bibinfo{volume}{113}
  (\bibinfo{year}{1983}) \bibinfo{pages}{58--64}.
\bibitem[{Yang et~al.(1986)Yang, Tucker, Cheng, and Adamson}]{Yang1986}
\bibinfo{author}{W.~Yang}, \bibinfo{author}{R.~Tucker},
  \bibinfo{author}{B.~Cheng}, \bibinfo{author}{R.~Adamson},
  \bibinfo{journal}{J. Nucl. Mater.} \bibinfo{volume}{138}
  (\bibinfo{year}{1986}) \bibinfo{pages}{185--195}.
\bibitem[{Bangaru(1985)}]{Bangaru1985}
\bibinfo{author}{N.~Bangaru}, \bibinfo{journal}{J. Nucl. Mater.}
  \bibinfo{volume}{131} (\bibinfo{year}{1985}) \bibinfo{pages}{280--290}.
\bibitem[{Yang et~al.(1987)Yang, Yu, and Chen}]{Yang1987}
\bibinfo{author}{T.~Y. Yang}, \bibinfo{author}{G.~P. Yu},
  \bibinfo{author}{L.~J. Chen}, \bibinfo{journal}{J. Nucl. Mater.}
  \bibinfo{volume}{150} (\bibinfo{year}{1987}) \bibinfo{pages}{67--77}.
\bibitem[{Vitikainen and Nenonen(1978)}]{Vitikainen1978}
\bibinfo{author}{E.~Vitikainen}, \bibinfo{author}{P.~Nenonen},
  \bibinfo{journal}{J. Nucl. Mater.} \bibinfo{volume}{78}
  (\bibinfo{year}{1978}) \bibinfo{pages}{362--373}.
\bibitem[{Versaci and Ipohorski(1979)}]{Versaci1979}
\bibinfo{author}{R.~Versaci}, \bibinfo{author}{M.~Ipohorski},
  \bibinfo{journal}{J. Nucl. Mater.} \bibinfo{volume}{80}
  (\bibinfo{year}{1979}) \bibinfo{pages}{180--183}.
\bibitem[{Krasevec(1981)}]{Krasevec1981}
\bibinfo{author}{V.~Krasevec}, \bibinfo{journal}{J. Nucl. Mater.}
  \bibinfo{volume}{98} (\bibinfo{year}{1981}) \bibinfo{pages}{235--237}.
\bibitem[{Kuwae et~al.(1983)Kuwae, Sato, Higashinakagawa, Kawashima, and
  Nakamura}]{Kuwae1983a}
\bibinfo{author}{R.~Kuwae}, \bibinfo{author}{K.~Sato},
  \bibinfo{author}{E.~Higashinakagawa}, \bibinfo{author}{J.~Kawashima},
  \bibinfo{author}{S.~Nakamura}, \bibinfo{journal}{J. Nucl. Mater.}
  \bibinfo{volume}{119} (\bibinfo{year}{1983}) \bibinfo{pages}{229--239}.
\bibitem[{{Van Der Sande} and Bement(1974)}]{Vandersande1974}
\bibinfo{author}{J.~{Van Der Sande}}, \bibinfo{author}{A.~Bement},
  \bibinfo{journal}{J. Nucl. Mater.} \bibinfo{volume}{52}
  (\bibinfo{year}{1974}) \bibinfo{pages}{115--118}.
\bibitem[{Versaci and Ipohorski(1983)}]{Versaci1983}
\bibinfo{author}{R.~Versaci}, \bibinfo{author}{M.~Ipohorski},
  \bibinfo{journal}{J. Nucl. Mater.} \bibinfo{volume}{116}
  (\bibinfo{year}{1983}) \bibinfo{pages}{321--323}.
\bibitem[{Leli\`{e}vre et~al.(2002)Leli\`{e}vre, Fruchart, Convertc, and
  Lef\`{e}vre-Joud}]{Lelievre2002}
\bibinfo{author}{G.~Leli\`{e}vre}, \bibinfo{author}{D.~Fruchart},
  \bibinfo{author}{P.~Convertc}, \bibinfo{author}{F.~Lef\`{e}vre-Joud},
  \bibinfo{journal}{J. Alloys Compd.} \bibinfo{volume}{347}
  (\bibinfo{year}{2002}) \bibinfo{pages}{288--294}.
\bibitem[{Okamoto(2007)}]{Okamoto2007}
\bibinfo{author}{H.~Okamoto}, \bibinfo{journal}{J. Phase Equilib. Diffus.}
  \bibinfo{volume}{28} (\bibinfo{year}{2007}) \bibinfo{pages}{409--409}.
\bibitem[{Clark et~al.(2005)Clark, Segall, Pickard, Hasnip, Probert, Refson,
  and Payne}]{Clark2005}
\bibinfo{author}{S.~J. Clark}, \bibinfo{author}{M.~D. Segall},
  \bibinfo{author}{C.~J. Pickard}, \bibinfo{author}{P.~J. Hasnip},
  \bibinfo{author}{M.~I.~J. Probert}, \bibinfo{author}{K.~Refson},
  \bibinfo{author}{M.~C. Payne}, \bibinfo{journal}{Zeitschrift f\"{u}r
  Kristallographie} \bibinfo{volume}{220} (\bibinfo{year}{2005})
  \bibinfo{pages}{567--570}.
\bibitem[{Domain et~al.(2002)Domain, Besson, and Legris}]{Domain2002a}
\bibinfo{author}{C.~Domain}, \bibinfo{author}{R.~Besson},
  \bibinfo{author}{A.~Legris}, \bibinfo{journal}{Acta Mater.}
  \bibinfo{volume}{50} (\bibinfo{year}{2002}) \bibinfo{pages}{3513--3526}.
\bibitem[{Perdew et~al.(1996)Perdew, Burke, and Ernzerhof}]{Perdew1996}
\bibinfo{author}{J.~P. Perdew}, \bibinfo{author}{K.~Burke},
  \bibinfo{author}{M.~Ernzerhof}, \bibinfo{journal}{Phys. Rev. Lett.}
  \bibinfo{volume}{77} (\bibinfo{year}{1996}) \bibinfo{pages}{3865--3868}.
\bibitem[{Monkhorst and Pack(1976)}]{Monkhorst1976}
\bibinfo{author}{H.~J. Monkhorst}, \bibinfo{author}{J.~D. Pack},
  \bibinfo{journal}{Phys. Rev. B} \bibinfo{volume}{13} (\bibinfo{year}{1976})
  \bibinfo{pages}{5188--5192}.
\bibitem[{Methfessel and Paxton(1989)}]{Methfessel1989}
\bibinfo{author}{M.~Methfessel}, \bibinfo{author}{A.~Paxton},
  \bibinfo{journal}{Phys. Rev. B} \bibinfo{volume}{40} (\bibinfo{year}{1989})
  \bibinfo{pages}{3616--3621}.
\bibitem[{Okamoto(2006)}]{Okamoto2006}
\bibinfo{author}{H.~Okamoto}, \bibinfo{journal}{J. Phase Equilib. Diffus.}
  \bibinfo{volume}{27} (\bibinfo{year}{2006}) \bibinfo{pages}{548--549}.
\bibitem[{Kearns(1967)}]{Kearns1967}
\bibinfo{author}{J.~Kearns}, \bibinfo{journal}{J. Nucl. Mater.}
  \bibinfo{volume}{22} (\bibinfo{year}{1967}) \bibinfo{pages}{11}.
\bibitem[{Lumley et~al.(2012)Lumley, Murphy, Burr, Grimes, Chard-Tuckey, and
  Wenman}]{Lumley2012}
\bibinfo{author}{S.~C. Lumley}, \bibinfo{author}{S.~T. Murphy},
  \bibinfo{author}{P.~A. Burr}, \bibinfo{author}{R.~W. Grimes},
  \bibinfo{author}{P.~R. Chard-Tuckey}, \bibinfo{author}{M.~R. Wenman},
  \bibinfo{title}{{The Stability of Alloying Agents in Zirconium}},
  \bibinfo{year}{2012}. \bibinfo{note}{Submitted to J. Nucl. Mater.}
\bibitem[{Lide(2010)}]{CRC90}
\bibinfo{author}{D.~R. Lide}, \bibinfo{title}{{Handbook of Chemistry and
  Physics, 90th edition}}, \bibinfo{publisher}{CRC}, \bibinfo{edition}{90}
  edition, \bibinfo{year}{2010}.
\bibitem[{Hahn(1993)}]{Hahn1993}
\bibinfo{editor}{T.~Hahn} (Ed.), \bibinfo{title}{{International Tables for
  Crystallography}}, volume~\bibinfo{volume}{A} of
  \textit{\bibinfo{series}{International Tables for Crystallography}},
  \bibinfo{publisher}{International Union of Crystallography},
  \bibinfo{address}{Chester, England}, \bibinfo{year}{1993}.
\bibitem[{Yamanaka et~al.(1995)Yamanaka, Higuchi, and Miyake}]{Yamanaka1995}
\bibinfo{author}{S.~Yamanaka}, \bibinfo{author}{K.~Higuchi},
  \bibinfo{author}{M.~Miyake}, \bibinfo{journal}{Journal of Alloys and
  Compounds} \bibinfo{volume}{231} (\bibinfo{year}{1995})
  \bibinfo{pages}{503--507}.
\bibitem[{Ells and Mcquillan(1956)}]{Ells1956}
\bibinfo{author}{C.~E. Ells}, \bibinfo{author}{A.~D. Mcquillan},
  \bibinfo{journal}{J. Inst. Met.} \bibinfo{volume}{85} (\bibinfo{year}{1956})
  \bibinfo{pages}{89--96}.
\bibitem[{Mallett and Albrecht(1957)}]{Mallett1957}
\bibinfo{author}{M.~W. Mallett}, \bibinfo{author}{W.~M. Albrecht},
  \bibinfo{journal}{Journal of The Electrochemical Society}
  \bibinfo{volume}{104} (\bibinfo{year}{1957}) \bibinfo{pages}{142}.
\bibitem[{Lim et~al.(2003)Lim, Hong, and Lee}]{Lim2003}
\bibinfo{author}{B.~Lim}, \bibinfo{author}{H.~Hong}, \bibinfo{author}{K.~Lee},
  \bibinfo{journal}{Journal of nuclear materials} \bibinfo{volume}{312}
  (\bibinfo{year}{2003}) \bibinfo{pages}{134--140}.

\end{thebibliography}








\begin{table}[p]
\centering
\caption{ \label{tab:SuperCells} Supercells adopted for each of the phases, in terms of the number ({\it p}, {\it q}, {\it r}) of repeats of the basic crystallographic unit cell parameters {\bf \emph{a}}, {\bf \emph{b}} and {\bf \emph{c}}, the total number of non-hydrogen atoms {\it N}, the shortest distance {\it d} between defect images in \si{\angstrom}, as well as the equivalent hydrogen concentration [H] in {\small{\it wt}~ppm}.}
\begin{tabular}{l l c@{ }c@{ }c r r@{.}l r}
\toprule
Phase	&Space	&$p$&$q$&$r$	&\multicolumn{1}{c}{{\it N}}	&\multicolumn{2}{c}{\it{d}}	& \multicolumn{1}{c}{[H]}\\
	&Group	&	&	&	&					&\multicolumn{2}{c}{}		&				\\
\midrule
$\alpha$-Zr	&			$\mathrm{P}6_{3}/mmc$	&5&5&3&	150&	15&6	&	73.7\\
$\beta$-Zr	&			$\mathrm{I}m\bar3m$	&4&4&4&	128&	14&3	&	86.3\\
\midrule
$\alpha$-ZrCr$_2$ {\it C15} &	$\mathrm{F}d\bar3m$	&2&2&2&	192&	14&3	&	80.7\\
$\beta$-ZrCr$_2$ {\it C36} &	$\mathrm{P}6_{3}/mmc$	&2&2&1&	96&	10&1	&	161.3\\
$\gamma$-ZrCr$_2$ {\it C14}&	$\mathrm{P}6_{3}/mmc$	&2&2&2&	96&	10&2	&	161.3\\
\midrule
ZrFe$_2$  {\it C15} &		$\mathrm{F}d\bar3m$	&2&2&2&	192&	14&1	&	77.6\\
ZrFe$_2$  {\it C36} &		$\mathrm{P}6_{3}/mmc$	&2&2&1&	96&	10&0	&	155.2\\
ZrFe$_2$  {\it C14} &		$\mathrm{P}6_{3}/mmc$	&2&2&2&	96&	10&0	&	155.2\\
Zr$_2$Fe &				$\mathrm{I}4/mcm$	&2&2&2&	96&	  9&8	&	132.2\\
\midrule
Zr$_2$Ni &				$\mathrm{I}4/mcm$	&2&2&2&	96&	10&5	&	130.6\\
\bottomrule
\end{tabular}
\end{table}

\begin{table}[p]
\centering
\caption{\label{tab:Esol-Zr} Enthalpy of solution of H in Zr in \si{eV} The value by Domain {\it et al.}\ is from DFT simulations, the others are all experimentally derived from pressure-concentration-temperature studies.}
\begin{tabular}{l r@{.}l S}
\toprule
Ref.				&\multicolumn{2}{c}{$\alpha$-Zr}	&\multicolumn{1}{c}{$\beta$-Zr} \\
\midrule
Current					&	-0&464			&-0.619\\
Domain\cite{Domain2002a}		&	-0&600			&\\
Yamanaka\cite{Yamanaka1995}	&	-0&424			&-0.455\\
Ells\cite{Ells1956}			&	-0&473			&-0.688\\
Mallett\cite{Mallett1957}		&	-0&373			&-0.077\\
Kearn	s\cite{Kearns1967}		&	-0&513$\pm$0.09	&\\
\bottomrule
\end{tabular}•
\end{table}•

\begin{table}[p]
\centering
\caption{\label{tab:Esol_summary} Enthalpy of solution for an interstitial H in Zr and its intermetallic phases. Positions are expressed following Wyckoff notation. $\Delta E_{sol}^H$ is the difference in enthalpy of solution of H in the given intermetallic, compared to the tetrahedral site in $\alpha$-Zr.}
\begin{tabular}{l l  r @{.} l r @{.} l r}
\toprule
 Phase &  Position & \multicolumn{2}{c}{$E_{sol}^H$ (eV)} & \multicolumn{2}{c}{$\Delta E_{sol}^H$(eV)} \\ \midrule
$\alpha$-Zr			&tet-{\it 4f} 	& -0&46	& \multicolumn{2}{c}{--}	\\
				&oct-{\it 2a}		& -0&38	&  0&09			\\
$\beta$-Zr			&tet-{\it 12d}		& -0&62	& -0&16			\\
				&oct-{\it 6b}		& -0&41	&  0&05			\\
\midrule
$\alpha$-ZrCr$_2$ ({\it C15})	&tet-{\it 96g}	& -0&30	&  0&16		  	\\
$\beta$-ZrCr$_2$ ({\it C36})	&tet-{\it 6h}	& -0&28	&  0&19		  	\\
$\gamma$-ZrCr$_2$ ({\it C14})	&tet-{\it 6h}	& -0&36	&  0&10		  	\\
ZrFe$_2$	({\it C15})	&tet-{\it 96g}		&  0&04	&  0&50	 	 	\\
ZrFe$_2$	({\it C36})	&tet-{\it 6h}		&  0&04	&  0&50  			\\
ZrFe$_2$	({\it C14})	&tet-{\it 6h}		&  0&03	&  0&50	  		\\
Zr$_2$Fe			&tet-{\it 16l}		& -0&45	&  0&01			\\
Zr$_2$Ni			&tet-{\it 16l}		& -0&67	& -0&20		  	\\
\bottomrule
\end{tabular}
\end{table}

•

\end{document}